# Time Allocation and Long-Term Scheduling of ESO Telescopes at La Silla Paranal Observatory

M. Rejkuba*a, O. R. Hainaut**a, T. Bierwirtha, M. Pruemma, A. Weissa,b
aEuropean Southern Observatory, Karl-Schwarzschild-Strasse 2, 85748 Garching b. München, Germany; bFEV etamax GmbH

**ABSTRACT**

The preparation of the long-term telescope schedule follows the submission and scientific review of new proposals. At the European Southern Observatory (ESO) this process entails scheduling the scientific proposals according to their scientific merit and available observing resources, as well as scheduling technical, maintenance, and commissioning activities for all operational telescopes of the La Silla Paranal Observatory. After the recent overhaul of the phase 1 proposal submission software, ESO started the development of a new telescopes time allocation and scheduling tool. The new tool makes the scheduling process more efficient while optimising the use of observing facilities. Besides scheduling activities and allocating time according to scientific merit, available resources, operational and programmatic needs, the tool will enable simultaneous scheduling of multiple telescopes to appropriately account for dependencies between them. The implementation of this new Time Allocation tool opens the possibility for dynamic re-scheduling of the telescopes, which is a pre-requisite to implementing a yearly "Call for Proposals" along with the "Fast Track Channel" at ESO.

## 1. INTRODUCTION

The European Southern Observatory (ESO) is an intergovernmental organization with 16 member countries, a partnership with Australia, and Chile, the host country, where the observatories are located. ESO's mission includes building and operating ground-based telescopes in Chile, enabling astronomers from all over the world to carry out cutting-edge scientific observational projects.

ESO's observatories in Chile include the La Silla Paranal Observatory (LPO) and the partnership in ALMA. Currently, within the LPO, ESO offers observing time and operates two 4m-class telescopes, NTT and 3.6m, on La Silla. On Cerro Paranal, it operates the four 8m Unit Telescopes (UTs) of the Very Large Telescope (VLT), the VLT Interferometer (VLTI) that combines light either from the four UTs or uses four 2m-class Auxiliary Telescopes (ATs), as well as VISTA survey telescope. Furthermore, on Cerro Armazones ESO is constructing the 39m Extremely Large Telescope (ELT), whose operations will be integrated within the LPO end-to-end dataflow system (Hainaut et al. 2018 [1]).

The LPO telescopes are equipped with a large and diverse set of instruments. In La Silla, the NTT currently has EFOSC2 and visitor instruments ULTRACAM and AstraLux and will soon be equipped with the SoXS instrument focusing on transients and variable objects science cases, while the 3.6m equipped with HARPS and NIRPS high-resolution spectrographs is one of the key exoplanets' exploration facilities from the ground. On Paranal, each of the UTs can be equipped with a set of instruments that occupy its two Nasmyth foci, a Cassegrain, and Coudé foci. The different VLT instruments offer optical and near-IR imaging, and spectroscopy with a broad range of wavelength coverage from the atmospheric cutoff in ultraviolet to mid-infrared wavelengths as well as offering different spectral resolutions. Several instruments offer polarimetric imaging and/or polarimetric spectroscopic modes and some are also equipped with coronagraphs for high-contrast observations. Several instruments are furthermore equipped with adaptive optics modules for high spatial resolution observations. In the Incoherent Combined Coudé Focus (ICCF) of the VLT the light can be conducted either from a single UT or combined from all 4 UTs together for observations with the ESPRESSO instrument which enables highly stable and very high-resolution spectroscopic observations. The VLTI interferometer is equipped with 3 instruments which use either the four 8m UTs or the four 1.8m ATs together. The ATs can be moved to different positions on the observing platform, thus forming different combinations of baselines that can be used for extremely high-resolution interferometric imaging as well as spectroscopic observations.

* mrejkuba@eso.org
** ohainaut@eso.org

The observing modes offered at the LPO telescopes include the Visitor Mode (VM), where observers travel to the telescope, and the Service Mode (SM), where observations are taken by ESO staff according to scientific priority and prevailing atmospheric conditions. In 2014 the so-called `designated Visitor Mode' (dVM) was introduced (Marteau et al. 2016 [2]). This enabled remote interactions for short observing runs, which for scheduling or scientific reasons required fixed VM-like slots, but were logistically complex, non-sustainable, or inefficient for a classical VM[1]. More details about the usage and advantages offered by these different observing modes are described in Rejkuba et al. (2018 [3]).

The fleet of LPO telescopes and their associated instruments are operated within the integrated end-to-end operation process developed for the VLT in the 1990-ies (Quinn et al., 1998 [4]; 2002 [5]). Similar operations models are adopted by several observatories preparing for the era of Extremely Large Telescopes (Hainaut et al. 2022 [6]).

For over 60 years ESO has invited the astronomical community to submit observing proposals twice per year and observations have been scheduled accordingly in 6-month long semesters. In the 70-ies and 80-ies, the proposals were sent by post first to ESO, and then to the reviewers. The number of proposals was much lower than today, and their scheduling was done manually, blocking the time on the calendar. This "traditional" scheduling was revised with the introduction of Service Mode first tested on the NTT and then implemented as part of the VLT end-to-end integrated operations when the first UT started operating in 1998. An electronic submission of latex-based proposals stored relevant information in the operational databases opening a possibility for a software-based telescope scheduling tool. The software scheduler called 'Time allocation Tool' (TaToo) was designed in 2003 in collaboration with an external company (Alves et al. 2005 [7]).

Telescope time allocation is a complex process whose primary objective is to maximize the scientific productivity of the observatory by allocating time according to the scientific merit of the proposals (Patat & Hussain 2013 [8]). Over the past couple of decades, the speed of discovery has increased, and instruments are both more powerful and more complex, some of them requiring the parallel scheduling of multiple telescopes (e.g. ESPRESSO, VLTI), as opposed to the one-telescope-at-a-time concept at the base of TaToo. The number of proposals that ESO receives has also increased, stabilizing currently at slightly more than 900 proposals per 6-monthly call. Together with further Directors Discretionary Time (DDT) proposals the LPO currently receives the order of 2000 proposals per year that are reviewed and considered for scheduling. Different types of observing programmes have been introduced, such as multi-semester Large Programmes, Public Surveys, and Monitoring Programmes, and for transients and other urgent observations there are also Target of Opportunity and Rapid Response Mode types of observing runs. Furthermore, coordinated proposals are now solicited for joint observations on the VLT/VLTI and ALMA as well as for coordinated projects with the XMM, and ESO plans to change the frequency of the regular calls for proposals to one main yearly Call for Proposals introducing at the same time the Fast Track Channel (FTC), following the recommendation by the Time Allocation Working Group (Patat, 2018 [8]).

As part of the evolution of the LPO's end-to-end operations (Hainaut et al. 2018 [1]), ESO has evolved its Phase 2 observations preparation, short-term scheduling and execution tools (Bierwirth et al. 2010 [10]; Beccari et al. 2022 [11]), deployed a new Phase 1 proposals submission system (Primas et al. 2022 [12]), and is now in the process of upgrading its Time Allocation system. Recently, a new time allocation and scheduling operations team has been established within the User Support Department (USD) and a new telescope Time Allocation tool (TA2) has been designed. The first version of TA2 has been successfully used to produce the ESO Period 113 schedule at the beginning of January 2024. This paper describes the new LPO long-term scheduling operations[2], software tool architecture and functionalities along with the development plans.

---

[1] The establishment of dVM also enabled efficient resumption of observations as soon as the observatories could be partially re-opened in the 2020 pandemic.

[2] The short-term scheduling evaluates which of the accepted observations to execute in the next night or next hour(s). The long-term scheduling operation evaluates which proposed and highly ranked observing runs can fit within the available telescope observing time leading to a recommendation for proposals acceptance and is part of the observatory operations planning over a period of several months.

## 2. SCHEDULING APPROACH

The time allocation and scheduling tool (TA2) requires the following input for scheduling:
- Observing proposals, with associated science runs (defining the instrumental set-up and atmospheric, moon, and scheduling constraints) and observations (with target and requested telescope time), defined within the Phase 1 tool, P1
- Ongoing commitments to observing proposals
- Technical Time Request (TTR) proposals, with associated TTR runs and activities
- Scheduling configuration parameters (assignment of instruments to telescopes, weather condition statistics, etc).

The science runs with their associated observations are automatically imported from the Phase 1 database into the TA2 global database. The TTR runs are included in the TA2 global database after their approval and assignment of a scheduling priority by the Observatory director. The scheduling configuration parameters are set in the tool by the staff in charge of schedule preparation. The scheduling input is described further in subsections 2.1– 2.3.

The schedule preparation starts with the definition of a Long-Term Schedule (LTS) container within which the scheduling input is elaborated to produce the list of runs that are either allocated time (traditional SM runs), or get fixed nights scheduled on a calendar (VM, dVM, SM runs with stringent time constraints, and TTR runs). An LTS has a validity period, a filling profile, and a list of available telescope setups. The validity period defines the time range over which the telescope time must be allocated; the filling profile enables the production of schedules that are filled in some months and less full in others and is a provision for the future scheduling of the yearly call with the FTC proposals submission. The relevant science and TTR runs whose validity intersect the LTS validity period and that require telescope setups considered in the LTS are automatically imported from the global TA2 database into the LTS.

Several LTSs can be defined and worked upon during the schedule preparation. They are produced by cloning an existing LTS or by creating a new LTS from scratch. Different clones of an LTS can be used to test the effect of different scheduling assumptions, such as for example, different priorities, or the impact of a decision when there are conflicting requirements, and they enable independent work by different staff on the preparation of the schedule.

Upon import of the runs, the so-called astro-processing is carried out (see 2.4 for more details). The result of the astro-processing is one or more sub-runs for each SM/VM/dVM or TTR run. The possible splitting of a run into sub-runs is done according to scientific requirements, which are defined through time constraints associated with the run, or by the fact that not all targets can be observed during a single VM/dVM sub-run. The scheduling algorithm will either accept or reject the entire sub-run according to priority and available time.

Curation of the scheduling input (runs, targets, observations, and their parameters) is possible within the LTS browser and is only accessible to the staff with corresponding privileges. The rationale for and some examples of the necessary editing of the scheduling input are described in subsection 2.5.

Once the scheduling input is prepared, the scheduler can be invoked requesting to schedule one or several telescopes. Currently, the scheduler is running independently for each requested telescope. Due to this, the schedule preparation must be performed over several iterations resolving conflicts between different requirements: to analyse where the top-ranked time-critical observations must be placed, to define the scheduling requirements for the Guaranteed Time Observations proposals, to define the time intervals for the VLTI observations requesting UT baselines, and to decide where to schedule ESPRESSO-4UT runs that are ranked highly enough to be accepted, because they utilise the time on all 4 UTs of the VLT simultaneously. We expect to implement a version of the scheduler capable of processing multiple telescopes simultaneously in an upcoming version of the tool.

After the schedule computation is completed and the person in charge of the scheduling determines that the schedule is ready to be published, the LTS is reviewed and approved by the Director, then declared as 'master LTS', and its content, all the edited runs and their properties are copied into the global TA2 database overwriting the existing content. The TA2 global database is therefore the authoritative repository of run allocation and scheduling information. The webpage https://www.eso.org/LPOschedule provides public access to the schedule timeline and the list of accepted, i.e. scheduled or allocated, science runs.

In the subsequent subsections we describe some of the key aspects relevant for the scheduling.

**2.1 LTS input: Phase 1 observing proposals and ongoing observation time commitments**

Each observing proposal prepared and submitted with the Phase 1 tool has one or more observing runs. Each run has a validity period that defines the interval of time for the run scheduling or time allocation. Each run defines the observing setup, atmospheric (sky transparency, turbulence, water vapour), Moon, and time constraints, and lists the observations with their targets, airmass constraint, and requested telescope time. As an output of the Distributed Peer Review and Observation Program Committee review process, each observing run is assigned a grade based on its scientific merit. Grades are on a scale of 1 to 5 where the top grade is 1, 3 is the lowest grade for runs to be considered for inclusion in a new schedule, and 5 is the worst grade resulting in immediate rejection. The relative priority corresponding to the programme type (Large, Monitoring, Calibration, Normal, Guaranteed Time Observation (GTO) and GTO Large) and run type (Normal, ToO, RRM) is reflected in the grade. Science observing runs and their associated observations are automatically imported from the validated proposals in the Phase 1 database to the scheduling global database after the closure of the call for proposals or upon submission of a new Director Discretionary Time (DDT) proposal. Information about the proposing team is not shown in the scheduling tool, as it is not strictly relevant for scheduling, and thus possible biases are minimised.

The ongoing scientific commitments include allocations for Large Programmes, Public Surveys, and Monitoring programmes approved in previous semesters. These are included as new runs with the top grade for scheduling in the new observing period. The relevant data for the scheduler are taken from the global database, as done for the newly submitted proposals.

ESO Optical/Infrared Telescopes Science Operations Policies [13] prescribes that the highest priority SM runs, allocated in A-rank class, that are not completed at the end of their originally allocated period (e.g. due to unexpectedly long periods of weather losses or due to technical issues), can get carried over for one more full visibility time interval of their targets. Therefore, an estimate of the expected carryover needs to be considered for these high priority science runs. In the past, as well as in the first release of the new scheduling tool, such carryovers are "simulated" by creating place-holding observing proposals with runs that have relevant combinations of instrument + observing constraints + time on target for the expected carryover observations. The input for place-holder proposals is derived by the USD, based on the pending Phase 2 observations in the queues and the available time in the long-term schedule.
The new scheduling tool will address, in an upcoming version, this time consuming and somewhat subjective preparation of expected carryover provision by directly including the pending observations into TA2 as input for the scheduler.

**2.2 LTS input: Observatory constraints and technical time requests**

The observatory constraints include engineering, maintenance, and commissioning activities that limit the time available for scientific observations. These are requested for inclusion in the schedule via the technical time request (TTR) runs. In the past, the inclusion of the TTR runs in the long-term schedule was done manually blocking the calendar slots and exchanging the information via Excel sheet between the LPO Director and the responsible for the preparation of the new schedule.

In the new scheduling tool, the TTR runs are pushed into the scheduling global database using an API between the two tools. The TTR tool is designed and maintained by the LPO. The TTR proposals are reviewed and approved by the Observatory director. The approved TTRs are assigned grades on the same scale as the science runs so that they can be ranked together and scheduled jointly. In this way, the top priority TTRs and science runs get scheduled or allocated first, and the lower priority TTRs do not occupy fixed slots that would prevent the inclusion of some higher priority science runs.

Examples of TTRs include (i) time requests for new instrument installation and commissioning at a telescope, which can have medium or high priority and may have some flexibility for scheduling depending on the status of the project and required systems, (ii) major or complex maintenance activities such as telescope mirror coatings, which have top priority and typically relatively little flexibility for scheduling due to complex operation and high impact on science, or (iii) routine preventive maintenance (e.g. maintenance of the closed cycle cooler of an infrared instrument) which can be scheduled with more flexibility in a way to minimize impact on scientific observations. Besides the planned TTRs for software, electronics, mechanical, and other engineering activities necessary to keep the observing systems at their high performance and reliability, there may be also urgent fixes requested on shorter notice – such TTR runs are equivalent to new DDT proposal runs (or in the future new Fast Track Channel science proposals), leading to an important requirement for the scheduling tool to be able to handle re-scheduling and publishing of the updated schedule.

**2.3 LTS input: scheduling tool configuration**

The configuration of the tool is also an important input for the schedule preparation. The basic configuration includes: (i) the association between instruments and telescopes (denoting if an instrument is active and on which telescope it is located[3]), (ii) SM calibration overhead for execution of standard calibration plan on each instrument[4], (iii) observatory site dependent distribution of expected atmospheric conditions relevant for the SM time allocation (i.e., the probability of realization of atmospheric turbulence (seeing and coherence time), sky transparency and precipitable water vapour), (iv) definition of binning for the observing constraints used by the scheduler, and (v) definition of requirements for scheduling that stem from the ESO Science Operations Policy [13]. The latter include the maximum fraction of time to be allocated to ToO observations, time to be reserved for future allocations (e.g. for new DDT proposals), allocation fraction for the host country proposals, and the definition of the SM rank classes A, B and C.

**2.4 Astro-preprocessing**

To facilitate and speed up the task of the scheduling engine, all astronomical calculations are performed in advance, and all the run level scheduling constraints are parsed and converted into simpler, abstracted constraints. The astronomical pre-processing produces its output as a set of JSON files.

For each night of the LTS validity period, the sidereal time (ST) at twilights and at midnight are computed, as well as the coordinates of the Moon, the illuminated fraction of the Moon, the time of moonrise and moonset. The contribution of the night to the total available time is computed for each bin of constraints, using the probability of realization of these constraints, and using the presence and brightness of the Moon.

The number and size of the constraint bins that define the scheduling resources parameter space are included as part of the basic configuration of the tool described above. The FLI is computed using Moon ephemerides. The other constraints are associated with a probability distribution (included as part of the scheduling tool configuration described above). The weather losses are accounted for using the "Cloud" sky transparency bin[5].

The usable ST range is computed for each observation of each run (rounded to minutes), using the target's coordinates, the requested maximum airmass, and, where relevant, the shadowing caused by other telescopes on the observatory platform, which typically affects VLTI observations with AT baselines. The requested constraints are converted to bin numbers.

For observations to be performed in Service Mode, this information is sufficient for statistical time allocation. Some runs, however, will have to be scheduled on the calendar. This is the case for VM/dVM runs, for most TTRs, as well as for SM runs with stringent time constraints. In these cases, the observations are packed in traditional runs of consecutive nights, either using the observability and duration of each observation with a basic greedy algorithm and/or using the time constraints. There can be cases where a single group of nights cannot accommodate the observations, either because the targets are spread over a wide range of right ascensions, or because the time constraints request separate (groups of) nights (e.g. for planetary transit observations, or monitoring). If needed, the run will be split into sub-runs to accommodate these constraints. In all cases, all possible groups of nights are considered and saved for the scheduler to choose from without having to recompute anything. In case a night is not fully occupied, the occupation is rounded up to the nearest 10$^{th}$ of the night, and this information is preserved for the scheduler. The remaining tenths of the night can be used for other runs, either scheduled on these fragments of nights or used by the pool of SM observations.

---

[3] The VLT instruments attached to Cassegrain or Nasmyth foci can be moved from one UT to another.

[4] SM observations from different observing runs can use the same standard calibrations, leading to more efficient use of available nighttime on the telescope. Depending on the fraction of time requested and allocated to an instrument in SM, the respective fraction used for standard calibrations needs to be subtracted. The fraction of time used for calibration plan observations for each instrument is estimated as an average calibration time used over the past 5 or more semesters and is tracked with the Dashboard for Operations Metrics at ESO (DOME; Primas et al. 2012 [14]).

[5] This is an approximation, because weather loss is not only due to cloudy conditions but can happen also with other sky transparency conditions, e.g. because of the strong wind, or seeing worse than 2 arcsec, or through a combination of poor weather conditions.

TTRs go through the same process, albeit with simplified rules. The constraints and observability of TTR's (sub)runs are expressed in the same way as the scientific runs so that the scheduler can deal with them in a similar way. Currently, the TTRs are exclusively scheduled in "Visitor Mode", but we envision that some maintenance tasks could be queued together with the science observations, and therefore allocated in SM.

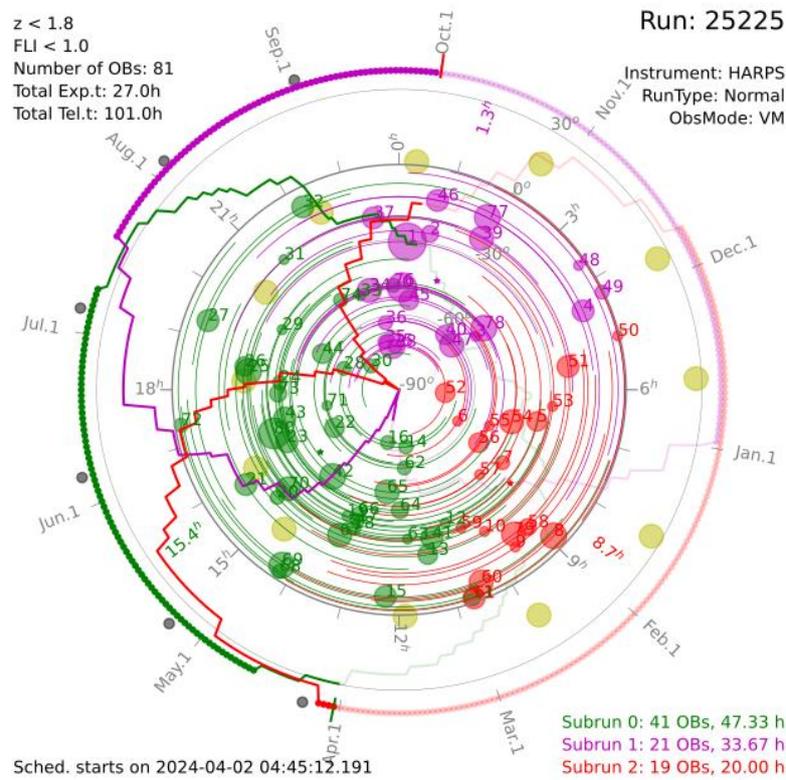

Figure 1. One of the output plots from the astro-pre-processing of a run, used to evaluate the requirements. In this polar view (azimuth is the right ascension or sidereal time, and the radius is the declination), each requested observation is a dot whose size indicates the duration of the observation, and whose "whiskers" mark the observability range. The dates on the rim are placed at their midnight sidereal time. The yellow dots mark the position of the full moons, and the black dots on the rim mark the time (and ST at midnight) of the new moons within the LTS validity range. In this case, the targets cannot be observed during a single contiguous set of nights, so the observations are spread over 3 sub-runs (green, purple, and red). The thick coloured dots on the rim mark the dates at which each sub-run could take place. The dates outside the validity range of the LTS are shown with muted colours.

A byproduct of astronomical preprocessing is a series of diagnostics, plots, and statistics that help in the curation process. Figure 1 illustrates the observation distribution of a VM run, which had to be split into three sub-runs. Figure 2 shows the observability of one sub-run, and for each possible starting date of the sub-run, indicated by coloured bars, it shows the duration of the group of nights needed to perform the observations (left y-axis) and the filling of the night occupation (right y-axis and thick red dots).

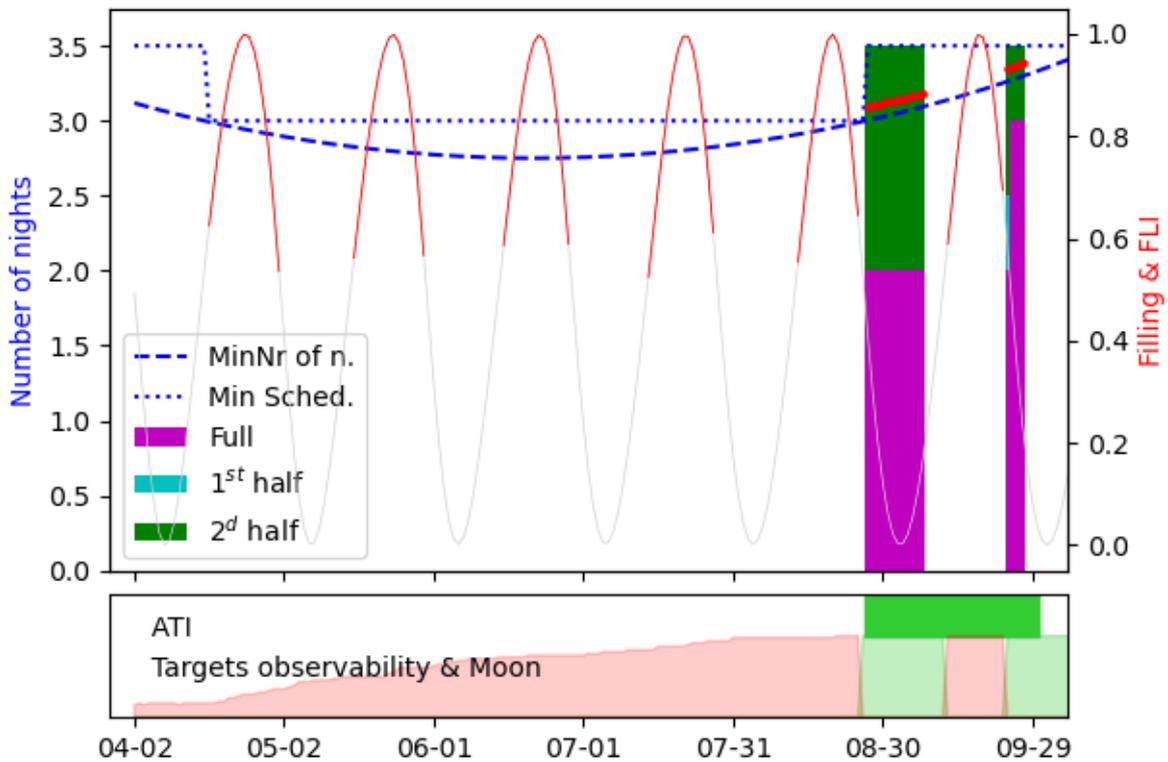

Figure 2. Another example of an output plot from the astro-pre-processing of another run, used to evaluate the requirements. For each night of the scheduling period (in x), the top plot shows the FLI (in grey line when matching the run's constraint, red otherwise), and the time needed to accommodate the requested observations (bars), in this case 3.5 nights, implemented as a combination of full nights (purple) and half nights (blue and green). These nights are not fully packed with observations: the filling factor is given as a thick red line, to be compared with the optimal filling (dashed blue line). The bottom plot shows the requested time interval (green bar), and the visibility of all the targets (red histogram when the targets are not all observable or when the moon is too bright, green otherwise). The scheduler will pick one of the possible solutions and implement it in the calendar.

**2.5 Curation of the scheduling input and preparation for scheduling**

The schedule elaboration requires many decisions and the resolution of conflicting requirements. Furthermore, some of the input received from Phase 1 proposals may need corrections or additions.

In the current implementation of TA2, the scheduler is invoked for each telescope independently. Therefore, for ESPRESSO-1UT observing runs (which can be executed on any of the four UTs), the runs must be edited to specify on which UT the run will be allocated time or scheduled. In the future, we plan to improve the scheduling algorithms enabling multiple-telescopes scheduling, which would open the possibility to optimize the allocation of ESPRESSO time on different UTs according to pressure as a function of RA and priority of other competing Phase 1 runs, rather than manually assigning the runs to a UT before running the schedule computation.

For VLTI the scheduling needs to consider grouping different observations that request the same VLTI baseline (combination of UTs or different locations of ATs) and therefore the requested baseline is an input along with the requested time for each observation. The requesters may specify alternative baselines in the Phase 1 proposal and if a run cannot be scheduled/allocated time with one baseline, the person preparing the schedule may decide to try modifying the input, assigning to the run the alternative (lower priority) baseline to try including the initially rejected run in the schedule.

Currently, the Phase 1 tool lacks some verification of the input, and we found some errors in the proposals. The most frequent errors are incompatibilities between the target declination, the requested airmass, and the duration of the observations, as well as incorrectly specified time constraints. Another case where input needs curation is for observations proposed for moving targets. In such cases, the coordinates of the proposed targets are either approximate or specified for one date. If the observations of moving targets are to be allocated in SM, it is better to ignore the targets' coordinates and specify the time intervals when the observations can be taken – then the astro-processing is triggered again to compute the relevant scheduling resources.

Other cases where the scheduling input must be modified include decisions to change the requested observing mode: for time-critical observations, such as exoplanet transits, that have only one or perhaps two possible nights per semester, there is no flexibility, which is the prerequisite for SM observations. Such runs are then converted to VM or dVM and the proposing team is informed about the decision and its rationale.

It is also possible to add a new run to an existing proposal. That is done by cloning the existing run and editing its parameters, targets, and observations. A typical example is if a separate SM pre-imaging run must be added to identify targets for the follow-up of multi-object spectroscopic observations, but the specification of a pre-imaging run was not included in the original proposal.

## 2.6 Scheduling

The scheduling tool is used to evaluate the possibility of including the observations requested within each sub-run. The science sub-runs are either included in the long-term schedule or rejected according to their scientific merit and availability of time at the telescope for the requested observations within constraints. Sub-runs requesting VM or dVM observing mode are scheduled on calendar dates, while SM sub-runs get time allocated on a statistical basis – the probability of completing the requested observations according to the targets' visibility and specified constraints. The time available for SM observations corresponds to the total time within the scheduling period minus the time scheduled for VM/dVM and TTR runs.

The scheduler first computes the LTS resources grid for the telescope. Each night contributes to the ST range between the evening and morning twilight as well as to one or two FLI bins according to the moon phase and its elevation.

The scheduler adds sub-runs one at a time to the schedule, in order of their priority. If no solution can be found fulfilling all scheduling constraints for the given sub-runs within a configurable maximum computation, the sub-run that was most recently tried to be included is rejected and the next sub-run is tried. This goes on until either all time is used or until all sub-runs are tried. The sub-runs are either fully allocated/scheduled or rejected.

The scheduler follows the following rules:

- Sub-runs are included in the order of priority that is given by their grade.

- Sub-runs scheduled on a calendar (VM/dVM, scheduled time-critical SM, TTR) must not overlap any other scheduled sub-run.

- VM/dVM sub-runs get scheduled on specific nights prioritizing the slots that have higher weight (the weight is assigned by astro-processing preferring denser packing of nights/shorter sub-runs).

- After successful scheduling of a VM/dVM sub-run, the respective ST and FLI bins are depleted (the scheduler knows the value of the ST and FLI for the calendar nights that are booked for the scheduled sub-run), while the other probabilistic constraints (sky transparency, PWV, and turbulence category) get depleted over all bins according to their probability of realization.

- After the successful allocation of SM sub-run the resources required by the respective observations are subtracted from the available resources starting from the loosest bin, and if not enough are remaining, the scheduler will subtract the resource(s) from the next more stringent bin(s). In other words, the SM observations requesting thin sky transparency may consume clear or even photometric constraints if not enough time within the thin constraint bin is remaining for their allocation.

All VLTI observations (requesting UTs or ATs) are considered as executed on a VLTI telescope. The scheduler is invoked with a dedicated algorithm that places all SM and VM/dVM observations on the calendar while keeping all observations using the same baseline in compact groups. For VM/dVM and scheduled SM runs, this result is kept as is, while the time slots used up by individual SM observations are labelled as generic "SM". In that way, the scheduler ensures that there is time for each and every SM observation and the flexibility of execution of SM observations is preserved. The groups of nights using ATs are labelled with their corresponding baselines, and the groups using UTs are "propagated" to the individual UTs to book the time on the individual telescopes. These groups of nights, when a given baseline is scheduled, can be adjusted and locked by the staff in charge of scheduling. In this way, after the initial guess is provided by the scheduler for the placement of the VLTI baselines, the scheduler can be invoked with the baseline groups fixed on the calendar to optimise the use of scheduling resources for SM and VM/dVM and TTR sub-runs within each set of baseline groups.

To change the VLTI baseline configuration, either ATs need to be moved to a different position on the telescope platform, or the VLTI needs to be configured to use UTs instead of ATs (or vice versa). At the end of each reconfiguration, which takes either one or two days, depending on the number of ATs that need to be relocated, one technical night is used for the system validation and calibration. Currently, VLTI-AT is offered with 4 different baselines (small, medium, large, and extended) and it can be used for snapshot, time-series, or imaging observations. For VLTI imaging, which is only offered in SM, it is sometimes necessary to have all 4 different VLTI-AT baselines within one month or less[6]. That is ensured by configuring the so-called Imaging Slots (ISL), which are time intervals during which only sub-runs requesting VLTI-AT baselines in SM are allocated time.

From the above it is clear that the VLTI scheduler has some additional rules due to interdependency between different telescope configuration setups.
- The VLTI scheduler must identify time intervals within which a given baseline configuration is scheduled, the so-called VLTI stretches, by grouping observations per baseline requested, or by receiving in input the set of VLTI stretches that are fixed.
- At any given night only one VLTI baseline is available, i.e. VLTI stretches must not overlap.
- At the beginning of each VLTI stretch, the scheduler needs to reserve one or two technical (relocation) nights.
- Imaging Slots are provided as input to the scheduler.
    - During the ISLs, only sub-runs requesting VLTI-AT configurations and SM observations are accepted
    - ISLs are only defined for VLTI-AT and therefore must not overlap VLTI-UT stretches
    - ISLs must overlap with at least one of the VLTI-AT baseline stretches

**2.7 Evaluation of the schedule, and iteration**

Once the scheduler has finished producing the schedule for one or more telescopes, the runs in the LTS are updated: the status of the sub-runs can be "allocated" or "not allocated". For the runs scheduled at fixed dates (VM, dVM, TTR, and SM with time constraints), the date of the first night and occupation of the scheduled nights are available.

Scheduled runs are displayed on a timeline (an example is given in Figure 3), and the database browser allows the TA2 operator with appropriate access privileges to check and evaluate each run. In particular, the rejection of highly ranked runs can be investigated. The person responsible for scheduling can check if the rejected run had targets in the highly oversubscribed RA range, check the proposal, and evaluate if it would be scientifically useful to reject a target in the oversubscribed range and try to schedule the remainder of the run. Such changes are exceptional and must be accompanied by appropriate scheduling comments for the proposing team. The tool offers the possibility to keep internal scheduling notes in the tool, as well as to add the public scheduling notes that become visible to the proposing team via the p1 tool interface after the schedule export. An allocated run can be "locked", ensuring that its status will be preserved in further iterations of the scheduling process.

A specific verification needs to be carried out for scheduled ESPRESSO-1UT runs. Given that the current scheduler runs on one telescope at a time, it is possible to have as a result two different ESPRESSO runs scheduled at the same time on two different UTs independently. This constitutes an overlap that can be seen by inspecting the ICCF telescope timeline. The ICCF shows all ESPRESSO scheduled sub-runs, both for 1UT and 4UT ESPRESSO observing setups. Currently, any overlap between ESPRESSO sub-runs needs to be fixed manually. In the future, we expect that the dependency between the telescopes can be handled by the multi-telescope scheduler algorithm.

---

[6] e.g. for image reconstruction of variable objects

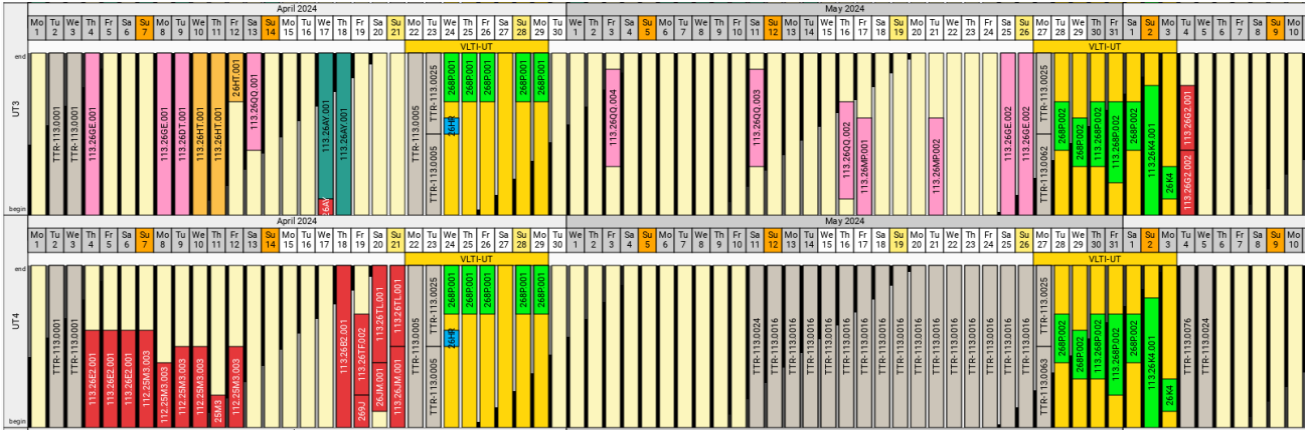

Figure 3. Example of a schedule timeline, for a few weeks, for two UTs. Each cell is a night (beginning of the night at the bottom of the cell, end at the top). Coloured bars indicate a scheduled run, the colour identifies the instrument, and the position the night occupation. Empty yellow areas mark time dedicated to SM observations. Note the VLTI observations (marked by a dark yellow bar), which occupy all UTs. Grey bars mark technical time requests. The thin white/grey/black margins of each cell indicate the brightness of the moon.

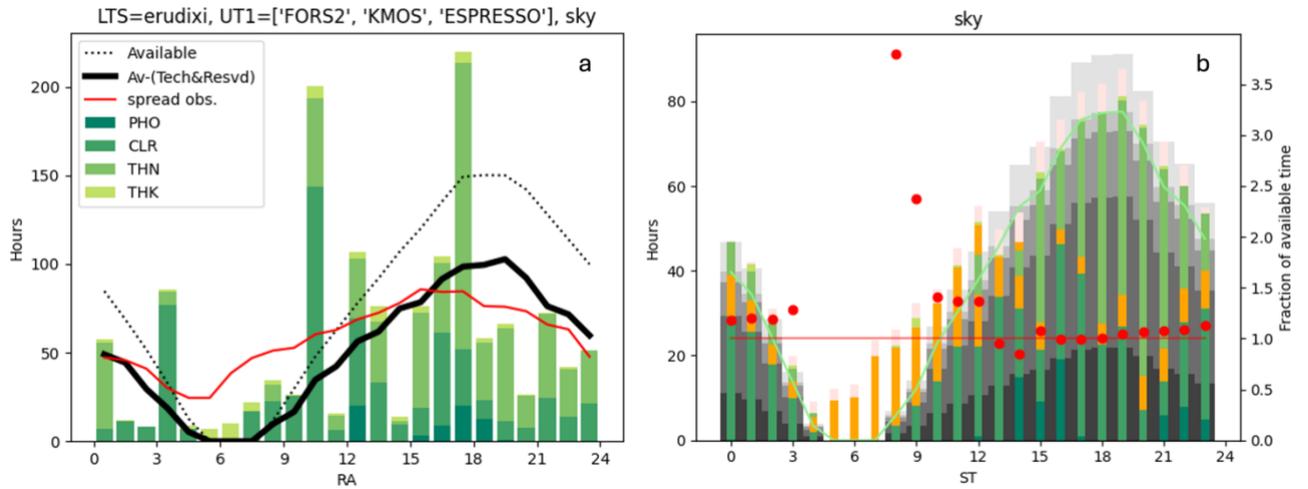

Figure 4. Two of the plots generated to evaluate a schedule. **Left (a):** The observing time requested, spread over twenty-four bins of sidereal time. The green bars show the request concentrated in the ST bin corresponding to the right ascension of the target (the colour indicates the constraint, here in terms of sky transparency ranging from photometric to thick cirri); the red line is the same, but spreading the observations over their full observable ST range; the thick black line shows the time available for observations after subtracting time for technical/engineering activities and reserved for the VLTI UT baseline. From this plot, the requested observations between 09 and 13h will overflow the available resources. **Right (b)**: Result of the schedule: the grey bars show the resources available for SM (also in terms of sky transparency, ranging from photometric in black, to cloud in light grey). The thin light green line is the time actually available for SM observations, excluding the "cloudy" conditions. The green bars indicate how these resources were used by the scheduler. The orange bars mark resources that were manually allocated, overruling the automatic scheduling. Light pink bars are runs that were added on top of the schedule and allocated as "filler" runs with loose constraints used to slightly over-populate the SM queues and provide flexibility in poor conditions (rank C runs). The red dots indicate the ratio between the allocated and available time, which should be slightly over 1. The two points with very high values are artifacts caused by the very small amount of time available in these bins.

A further check is carried out for the VM/dVM runs scheduled on UT4 that use laser assisted Adaptive Optics instrument(s) with respect to all other scheduled (VM/dVM/time-critical SM) runs on different UTs. The relative locations of all the targets from the laser (UT4) and non-laser (UT1/UT2/UT3) scheduled run are compared to evaluate if there are possible collisions that could negatively affect the non-laser observations due to the presence of sodium light as well as the addition of further Raman scattered features. The collision can happen when the two telescopes point to the same field at the same time, and when the laser beam from UT4 crosses the path between the target and the telescope for the non-laser observation. The latter case has a larger collision cross-section and is thus a more frequent cause of conflict.

A series of reports are produced to evaluate the results of the scheduler – some examples are displayed in Figure 4.

The allocated SM time is split equally between the A- and B-rank classes. The A-rank class runs have the highest priority for execution, while the B-rank class runs are allocated if graded below the runs that fill the top half of the available time and request observing constraints and ST ranges that are not occupied by the higher, A-ranked allocated runs.

The C-rank class is assigned to filler type runs, with fairly relaxed observing constraints – these runs are not formally allocated time, but are added on top of a fully filled long-term schedule to ensure that when the weather conditions do not permit observations of higher ranked (A- or B-rank) runs, but still allow to get some useful science data, the telescopes do not remain idle. The assignment of the C-rank class is done manually at the end of the schedule preparation. The filler type runs are pre-selected based on the requested constraints, but their suitability for C-rank class allocation as filler requires reading the proposals to check the science case and observing strategy. The expectation is that A-rank class runs are mostly completed, B-rank class runs to get as much completed as the weather and other higher priorities allow, while the C-rank class runs get only partial datasets.

**2.8 Publication of the schedule**

The previous steps are iterated to schedule all telescopes, possibly adjusting one of the schedules to accommodate constraints from another one (currently, we schedule each telescope independently; this will be addressed in an upcoming version of the tool), verifying the sanity and quality of the allocation. Eventually, a final Long-Term Schedule is produced. It is presented to the director of the observatory, the director for science, and the director general for approval. Once approval is secured, the LTS is declared "Master LTS". This action locks it to prevent further changes, and its content is copied into the global TA2 database, updating all the runs, their allocation status, scheduling information, comments, and notes.

The global TA2 database is the authoritative repository of scheduling information: it contains all runs and their up-to-date information from the latest master LTS. This global database can be accessed via public APIs (which filter out confidential information), and via a public web interface at http://www.eso.org/LPOschedule. A set of restricted APIs allows operation staff and tools to access part or all of the TA2 database, depending on their role's privileges.

For compatibility with tools still relying on the previous scheduling infrastructure, the information from the master LTS is also copied into the legacy scheduling database. These tools will be upgraded to access the information from the TA2 global database.

A new LTS can then be created, either "cloning" the master LTS (for minor changes such as schedule updates and inclusion of new DDT proposals) or creating a brand new one directly from the TA2 global database (to start working on a new proposal cycle). The new LTS is brought through the same process until it eventually becomes a new master LTS.

## 3. SOFTWARE ARCHITECTURE AND ADOPTED TECHNOLOGIES

In the following we describe the software architecture and adopted technologies. The overall software is built as a web-based client-server application with a rich user interface and a modular backend, consisting of the backend TA2 server and two modules for astronomical pre-processing and scheduling (Figure 5). Data is exchanged between the frontend and backend via well-defined REST-APIs. The backend submodules for astronomical calculations and the scheduling tasks are triggered via a message by the TA2 server, where the data exchange is again done via the server's REST-APIs.

Application data are persisted with a Couchbase Database, which is described in the next subsection.

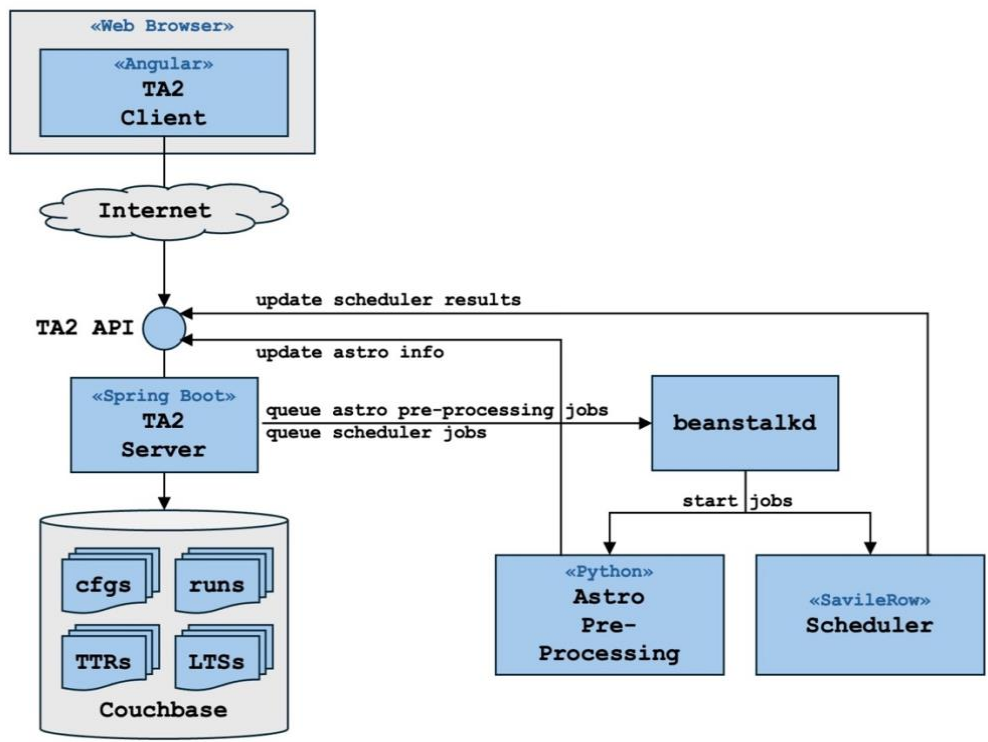

Figure 5. Logical SW architecture of TA2

**3.1 Database Persistence**

As opposed to a more traditional relational SQL database storing data in tables, we decided to use the NoSQL database Couchbase[7] for permanently storing data in the TA2 project. Couchbase is a document-oriented database that allows saving hierarchically structured, self-contained documents in JSON[8] format. Different *types* of documents are stored, such as basic configurations, science runs (with their contained targets and observations), TTRs, and the actual long-term schedules (LTS). For the preparation of the P113 schedule, we had the order of 1900 science runs, nearly 10,000 targets, and 11,000 observations. In comparison to storing such information in a relational DB using an object-relational mapping approach, we found both the runtime performance of Couchbase for storing and querying data and also the productivity of our SW engineers significantly improved.

**3.2 Server: Business Logic and APIs**

The domain logic for the scheduling approach (see Section 2) is implemented on the server side in terms of a Spring Boot[9] application using Groovy as the programming language and Spring Data Couchbase[10] as the interface to our Couchbase. This domain logic is exposed via an API with 50 plus REST endpoints. Since the API is used by fully empowered TA2 operators, privileged (but read-only) internal operations users, and external users all endpoints are secured with authentication via OIDC and role-based authorisation.

We follow a well-defined approach in which we first, in joint discussions that include astronomers and SW engineers, detail requirements into small, implementation-ready user stories with acceptance criteria, subsequently identify required new API endpoints or changes to existing ones, then formally specify these using OpenAPI[11] (Figure 6) and finally assign them to our SW engineers for review, implementation, and testing. The final acceptance is carried out by the project scientist.

---

[7] Couchbase https://www.couchbase.com
[8] The JavaScript Object Notation (JSON) Data Interchange Format https://datatracker.ietf.org/doc/html/rfc8259
[9] Spring Boot https://spring.io/projects/spring-boot
[10] Spring Data Couchbase https://spring.io/projects/spring-data-couchbase

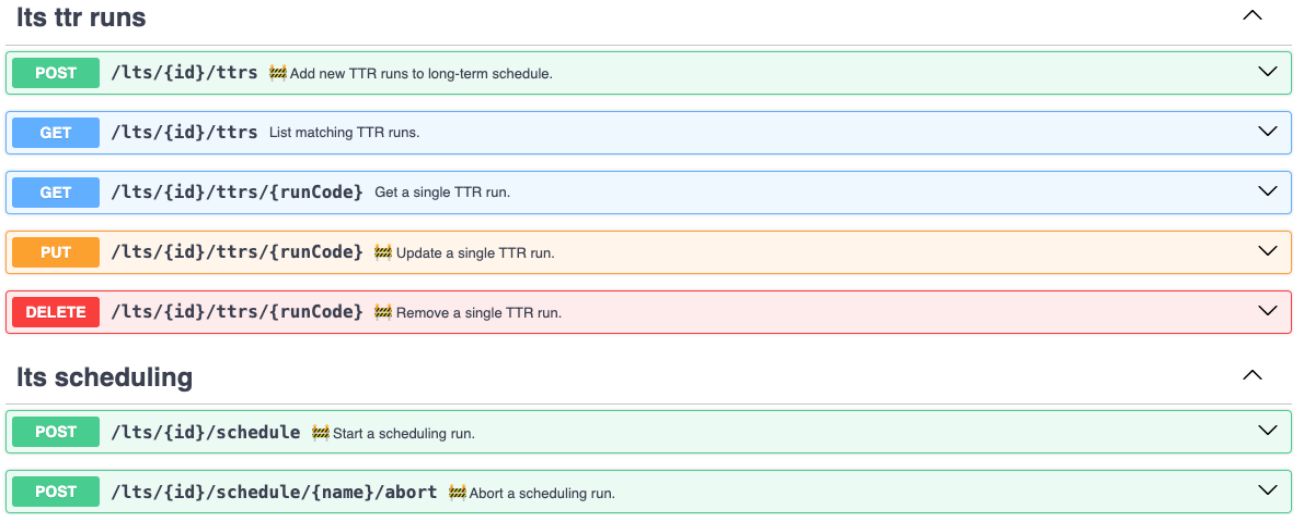

Figure 6. Formal specification of the TA2 API using OpenAPI.

### 3.3 Client: Business Logic and User Interface

The TA2 client is a modern "single-page application" implemented in the Typescript programming language using Google's well-established Angular[12] framework. It can be loaded and executed in any modern browser with a suitable JavaScript engine (Firefox, Chrome, Safari, etc.) and does not require any local installation. Bugfixes and releases of new features are instantly available to users upon browser refresh. To achieve a consistent user interface and to minimize required custom CSS styling, we use Angular Material[13] for most graphical user interface components. Depending on the user's assigned roles (e.g. anonymous or logged-in operator) the user interface offers a different set of functionalities.

The user interface of the client allows access to all client- and server-side business logic, such as editing the basic configuration, cloning/creating a new LTS, browsing the global DB of runs and TTRs and adding them to an LTS, curation of the runs and TTRs and their sub-runs within an LTS, manual overruling of the scheduler, starting and monitoring the scheduler, visualization of the scheduler results, browsing the changelog, declaration of the master LTS. The user interface also provides several client-side-only "convenience" features that can be realized without additional APIs, but make the operator's job much more efficient, such as selecting several runs/TTRs and carrying out multiple editing of their properties in one go. The client should not be mistaken as only providing a thin layer of graphical UI elements. It is a significant and complex application itself. We consider it of great importance to follow a well-defined, consistent approach to client-side state management from the start to ensure scalability and maintainability. We use the NgRx SignalStore[14] for that purpose.

### 3.4 Astro pre-processing and reporting Python modules

The astro pre-processing and the reporting are implemented as Python modules.

The astro pre-processing module is interfaced with the TA2 infrastructure, which calls it for each new or modified run, extracts the required information from the output JSON files, and stores the summary and evaluation plots as a web page accessible from the TA2 user interface.

The reporting module is called using a series of command line scripts that interface with the TA2 databases using APIs and produce HTML reports.

---

[11] OpenAPI https://www.openapis.org
[12] Angular Framework https://angular.io
[13] Angular Material https://material.angular.io
[14] NgRx Signal Store https://ngrx.io/guide/signals/signal-store

To ensure maintainability, both modules use only "standard" additional Python libraries (NumPy, matplotlib, astropy).

Since astro pre-processing may take between several seconds to a dozen minutes depending on the complexity of a given run, it is invoked on a dedicated server using the beanstalkd[15] work queue infrastructure, where jobs can be queued for asynchronous execution using the Java Beanstalk Client[16]. When finished, the respective job updates the LTS with the computed run-splitting information.

### 3.5 Scheduler

The part of the scheduler that allocates runs to fixed time intervals on the calendar (VM, dVM, TTR, and SM with time constraints) applies in-depth constraint programming techniques: Instead of imperatively coding a brute-force algorithm that finds a solution, the scheduling rules given by the astronomers in "plain English" are translated into formal mathematical constraints by defining variables, their domains (i.e. their allowed discrete values such as Boolean or integer range) and the relationships (i.e. constraints) between these variables that must be fulfilled to constitute a valid solution. The mathematical product of the number of allowed values for each variable defines the complete size of the solution space that must potentially be probed by the constraint solver. It is the responsibility of the constraint solver to search the entire solution space in an effective way to arrive at a valid solution quickly.

We express the constraints in the Essence Prime[17] constraint specification language. The constraints are then read by SavileRow[18], a "modelling assistant for Constraint Programming" and passed to a configurable underlying solver. Different solvers and solver technologies (constraint solvers, SAT, and SMT solvers) are used for different scheduling tasks. Depending on the size and complexity of the input, the applicable constraints, and the configured timeout of the solver, the scheduling of a telescope can take several hours and is therefore also executed in the beanstalkd infrastructure.

We are not aware of any Artificial Intelligence / Deep Learning approach that would work equally well for solving our problem because we do not have enough schedules to serve as training material. Furthermore, we note that the constraints on the existing schedules have changed over time.

### 3.6 Access control

Users of TA2 authenticate via logging into the ESO User Portal. Each user has a series of "roles" that define the access and privileges to the various operational tools and databases (Chavan et al. 2006 [15]). For TA2 currently two roles have been created: (i) a super user role with full read-write access and privileges to run the scheduler used by the scheduling team in USD and (ii) a role for the operations staff who need read-only access to most information and read-write access to limited information. Users without these roles, as well as anonymous users, have access to the public schedule information only.

## 4. PROJECT STATUS AND FUTURE PLANS

TA2 is under active development. Its first release at the end of 2023 already achieved nearly complete feature parity with the previous scheduling tool (TaToo) used for schedule preparation over the last 20 years and it included some new features. One of the important changes with respect to the old tool that facilitated greatly the schedule preparation was improved pre-processing of science runs based on user specifications of the requested targets with airmass and moon constraints, as well as time constraints. The old tool could not access detailed descriptions of targets and observations from the Phase 1 DB, and it disregarded the moon distance and target declination, requiring manual checks and fixing for most of the VM runs. Another improvement is the ability of the scheduler to group VLTI observations and propose the time slots for VLTI baseline stretches scheduling. A major new feature is the public graphical view of the schedule that displays the calendar with all scientific and technical runs for each night in the selected period, in addition to providing the list of accepted runs. The schedule is always up to date as it reads the allocation status from the underlying database.

---

[15] Beanstalk Work Queue https://beanstalkd.github.io
[16] Java Beanstalk Client https://github.com/RTykulsker/JavaBeanstalkClient
[17] Essence Prime Constraint Specification Language https://www.csplib.org/Languages/EssencePrime
[18] Savile Row Modelling Assistant for Constraint Programming https://www-users.york.ac.uk/peter.nightingale/savilerow

For the second TA2 release, which is currently in preparation for the ESO Period 114 schedule release, the major new functionality is the interface to the LPO TTR tool with the streamlined import of the TTR runs, which can then be prioritised and scheduled along with the science runs. The VLTI scheduler is also undergoing an update to enable editing and locking of the VLTI baseline stretches, which opens the possibility for statistical allocation of SM time on VLTI for each set of baseline stretches. Further workflows are implemented to streamline operations tasks and include communication support for the evaluation of runs that qualify for scheduling as SM fillers, as well as for technical assessment of scheduled runs.

The third release of the scheduling tool planned at the end of 2024, is expected to address the provision of ongoing scientific commitments. We plan to replace the Phase 1 observations with Phase 2 observation blocks for SM allocated runs. The Phase 2 observation blocks contain the most accurate description of observations and by removing those observations that have been completed, we can provide to the scheduler the complete and up-to-date list of pending SM commitments. The schedule will then be elaborated in such a way that the ongoing SM and planned VM/dVM and TTR runs are allocated time from now until the end of their validity. The Phase 1 proposals from the new Call for Proposals will be added with their individual validity range, which may include partial overlap with some of the ongoing SM runs. This will open the avenue for a more dynamic (re-)scheduling, which is a pre-requisite for the adoption of the yearly call for proposals along with the Fast Track Channel.

A final module in TA2 will add a workflow for handling target and setup change requests, which are currently received through a web interface in the Phase 2 tool with follow-up processing handled via email. Integrating the target and observing setup approval process in the scheduling tool will ensure consistent content of the schedule and provide further possibility to expose the approved new targets to the Phase 2 tool for the observation preparation in a fully consistent manner.

In parallel with the development of the new TA2 functionalities that will enable new and enhance the existing scheduling workflows, there are also further important improvements for the scheduler. For example, new rules need to be added for scheduling ESPRESSO instrument runs. The ESPRESSO-4UT runs scheduling is conceptually relatively simple: such runs occupy all 4UTs, must be scheduled in VM (i.e. occupy a (fraction of) calendar night), and are not compatible with any other scheduled run on a single UT or on VLTI with UT baseline. When scheduling ESPRESSO-1UT the scheduler can allocate/schedule ESPRESSO runs on only one UT within any given night (fraction), which implies that it needs to know the status of other UTs. Furthermore, any TTR or VM/dVM run scheduled on a single UT prevents the use of VLTI with UT baseline, but the VLTI with ATs can be available for scheduling on those nights. In other words, the telescopes' long-term schedule will need to be done simultaneously solving for all 4 UTs of the VLT as well as for the VLTI with UT and AT baselines. This will decrease the number of iterations necessary to produce a valid new long-term schedule and will open a possibility for further optimisation of time allocation according to scientific priority and pre-existing commitments on all telescopes.

## ACKNOWLEDGEMENTS

The development of the new scheduling tool would not have been possible without the collaborative spirit of colleagues from across ESO. We would like to thank all of them for their contributions. Nando Patat and Andreas Kaufer generously provided their experience as well as specific input for preparation and feedback on the first schedule prepared with the new scheduling tool in January 2024. Joe Anderson, Lowell Tacconi-Garman, Thomas Szeifert, Markus Wittkowski and Andrea Mehner contributed requirements specifications and use cases. Henri Boffin provided valuable feedback on the tool with suggestions for improvements to the existing and the addition of new features. We also benefitted from discussions about scheduling with Tereza Jerabkova, Elisabeth Hoppe, Steffen Mieske, Michael Sterzik, Xavier Haubois, and Antoine Mérand. Vicente Lizana and Stéphane Brillant worked on the TTR interface to TA2. Yves Jung and Jorge Grave developed the software components necessary for the schedule export to the old scheduling databases. Thanks are also due to Maksym Korpan, who joined the software development team in Q2 2024 and is contributing to the tool.